\documentclass[10pt,journal,final,letterpaper,twocolumn,twoside]{IEEEtran}

\usepackage{epsfig,amsfonts,amsmath,amssymb,verbatim}
\def\rR{{\mathbb R}}
\def\cC{{\mathbb C}}
\def\eE{{\mathbb E}}

\def\pP{{\mathbb P}}
\newcommand{\secref}[1]{Section~\ref{#1}}
\newcommand{\eqnref}[1]{(\ref{#1})}
\newcommand{\thmref}[1]{Theorem~\ref{#1}}
\newcommand{\lemref}[1]{Lemma~\ref{#1}}

\newcommand{\figref}[1]{Figure~\ref{#1}}
\newcommand{\prpref}[1]{Proposition~\ref{#1}}

\newtheorem{theorem}{Theorem}[section]

\newtheorem{lemma}{Lemma}[section]
\newtheorem{proposition}{Proposition}[section]
\newtheorem{corollary}{Corollary}[section]

\begin{document}

\title{Non--Parametric Field Estimation with Randomly Deployed,
Noisy, Binary Sensors$^{\text{\small 1}}$}

\author{
\authorblockN{Ye Wang and Prakash Ishwar}\\
\authorblockA{Department of Electrical and Computer Engineering\\
Boston University, Boston, MA \\
{\tt \{yw,pi\}@bu.edu}}}

\maketitle

\begin{abstract}
The reconstruction of a deterministic data field from
binary--quantized noisy observations of sensors which are randomly
deployed over the field domain is studied. The study focuses on the
extremes of lack of deterministic control in the sensor deployment,
lack of knowledge of the noise distribution, and lack of sensing
precision and reliability. Such adverse conditions are motivated by
possible real--world scenarios where a large collection of
low--cost, crudely manufactured sensors are mass--deployed in an
environment where little can be assumed about the ambient noise. A
simple estimator that reconstructs the entire data field from these
unreliable, binary--quantized, noisy observations is proposed.
Technical conditions for the almost sure and integrated mean squared
error (MSE) convergence of the estimate to the data field, as the
number of sensors tends to infinity, are derived and their
implications are discussed. For finite--dimensional,
bounded--variation, and Sobolev--differentiable function classes,
specific integrated MSE decay rates are derived. For the first and
third function classes these rates are found to be minimax order
optimal with respect to infinite precision sensing and known noise
distribution.
\end{abstract}


\section{Introduction}
\addtocounter{footnote}{+1} \footnotetext{This material is based
upon work supported by the US National Science Foundation (NSF)
under award (CAREER) CCF--0546598. Any opinions, findings, and
conclusions or recommendations expressed in this material are those
of the authors and do not necessarily reflect the views of the NSF.
A part of this work was presented at the 2007 International
Symposium on Information Theory (ISIT).}

In a recent paper~\cite{WangIshwarISIT07} we considered the problem
of reconstructing a bounded deterministic multidimensional data
field $f: [0,1]^p \rightarrow [a,-a],~0 < a < \infty$, from noisy
dithered binary--quantized observations collected by $n$ sensors
randomly deployed over the field domain. The random sensor
deployment model was based on uniform Monte Carlo sampling locations
where $n$ sensors are independently and identically distributed
(iid) uniformly over the field domain\footnote{The field domain
$[0,1]^p$ is used for clarity and ease of exposition. However, the
results can be generalized to compact subsets of ${\mathbb R}^p$.}
$[0,1]^p$. A simple estimator that reconstructs the entire data
field from these unreliable, binary--quantized, noisy observations
was proposed in \cite{WangIshwarISIT07} and an upper bound on the
integrated MSE of the estimator was derived. Using this bound, the
integrated MSE convergence of the estimator to the actual field as
the number of sensors $n \longrightarrow \infty$ was established.

In the present paper we expand and complete the development of
results in \cite{WangIshwarISIT07}: (i) In
Section~\ref{sec:perfResults} we expand the results of
\cite{WangIshwarISIT07} to general deployment distributions. We
establish a general upper bound to the integrated MSE which
highlights the interaction of the deployment distribution and the
orthonormal basis used for non-parametric field estimation
(Theorem~\ref{thm:MSEUpperBound}). (ii) We then derive sufficient
conditions on the deployment distribution, the orthonormal basis,
and the dimension of the field estimate which ensure the asymptotic
(as $n \longrightarrow \infty$) integrated MSE consistency of the
proposed estimator. Implications for desirable deployment
distributions are also discussed. (iii) In
Section~\ref{sec:ASConvResults} we comprehensively investigate the
asymptotic (as $n \longrightarrow \infty$) almost sure consistency
of the proposed estimator. The highlight of this section is
Theorem~\ref{thm:ErrorASConv} which provides an interesting set of
sufficient conditions on the deployment distribution, the
orthonormal basis, and the dimension of the field estimate which
ensures asymptotic almost sure consistency of the estimation error.
The implications of Theorem~\ref{thm:ErrorASConv} are explored in
detail through Proposition~\ref{prp:asAuxFcnsPrp} and
Corollary~\ref{thm:EstASConv} and are of independent interest.

For the finite--dimensional, bounded--variation, and
Sobolev--differentiable function classes, explicit achievable decay
rates for the integrated MSEs are provided in
Section~\ref{sec:MSEDecayRates}. Specifically, for fields that
belong to a finite--dimensional function space, the integrated MSE
decays as\footnote{Landau's asymptotic notation: $f(n) = O(g(n))
\Leftrightarrow \lim\sup_{n\rightarrow \infty}|f(n)/g(n)| < \infty$;
$f(n) = \Omega(g(n)) \Leftrightarrow g(n) = O(f(n))$; $f(n) =
\Theta(g(n)) \Leftrightarrow f(n) = O(g(n))\ \text{and}\ g(n) =
O(f(n))$.}  $O(1/n)$ (Corollary~\ref{thm:MSERateFiniteD}). For
fields of bounded--variation, the integrated MSE decays as
$O(1/\sqrt{n})$ (Corollary~\ref{thm:MSERateBV}).  For fields that
are $s$--Sobolev smooth (see \ref{sec:smoothFuncClass}), the
integrated MSE decays as $O(n^{\frac{-2s}{2s + 1}})$
(Corollary~\ref{thm:MSERateSmooth}).

One of the highlights of this work is that for multidimensional
fields living in rich function spaces, the minimax rate of
convergence, of the integrated MSE, even with randomly deployed
sensors, unknown noise statistics, and binary dithered scalar
quantization (a highly nonlinear operation), can match the minimax
rate of convergence with infinite--precision real--valued samples
and known noise statistics.

The application context of this work is distributed sensing and
coding for field reconstruction in wireless sensor networks as in
\cite{WangIshwarISIT07}. The focus is on the extremes of lack of
control in the sensor deployment, arbitrariness and lack of
knowledge of the noise distribution, and low--precision and
unreliability in the sensors. These adverse conditions are motivated
by possible real--world scenarios where a large collection of
low--cost, crudely manufactured sensors are mass--deployed in an
environment where little can be assumed about the ambient noise.
Each sensor measures a noisy sample of the field at its location
under iid zero--mean, bounded amplitude, additive noise.  The
statistical distribution of the noise is {\em unknown} to the
sensors and the fusion center, and the results in this paper hold
for {\em arbitrary} distributions satisfying these assumptions. Each
noisy sensor sample is quantized to a binary value by comparison
with a random threshold ($1$--bit dithered scalar quantization). The
binary--quantization models the extreme of low--precision
quantization. The random thresholds are assumed to be iid across the
sensors and uniformly distributed over the sample dynamic range,
modeling the extreme unreliability in the quantization across
sensors due to manufacturing process variations and environmental
conditions at different sensor locations. Such extreme modeling
assumptions are considered to demonstrate what is still achievable
under adverse conditions.

\begin{figure*}
\centering
\includegraphics[width=6.3in]{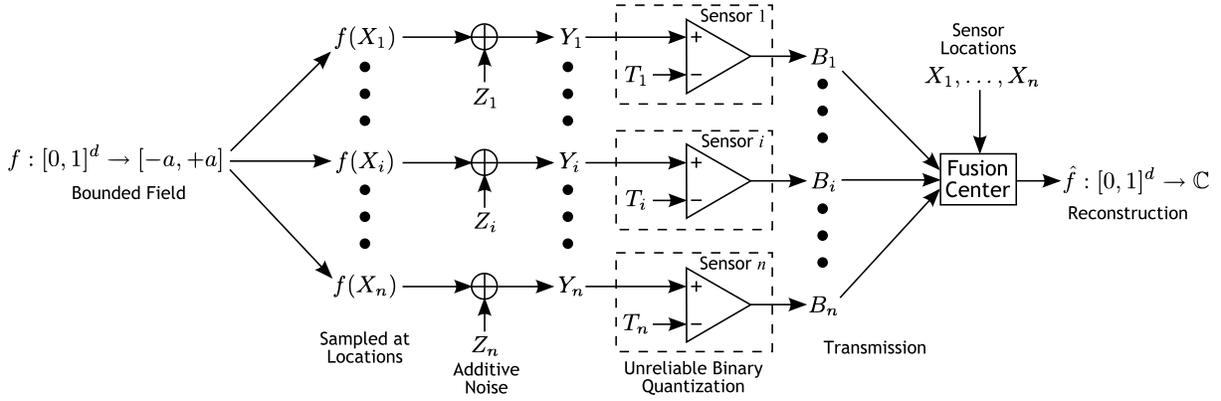}
\caption{{\bf Problem Setup:} \sl The field (in a single snapshot) is
sampled by $n$ sensors at their respective locations under additive
noise. Each sample is unreliably quantized to a binary value by a
comparison with a random threshold. These binary values are
transmitted to the fusion center which reconstructs the field.}
\label{fig:mainFig}
\end{figure*}

The communication channel issues are abstracted away by assuming
that the underlying sensor communication network is able to handle
the modest payload of transmitting one bit (the binary--quantized
observation) per sensor to the fusion center. The focus of this work
is on reconstructing a single time snapshot of the field at a fusion
center. The reconstruction of multiple time snapshots of the field
can also be accommodated within the framework of this work as in
\cite{WangMZIS-Allerton06-UDEoNFOb} but is omitted for clarity. In
fact, this can be achieved with time--sharing sensors, vanishing
per--sensor rate, and vanishing sensor location
``overheads''\footnote{Network overheads refer to additional bits of
information that must be attached to each message to identify the
point of origin of the message.} (see
\cite{WangMZIS-Allerton06-UDEoNFOb}). It is also assumed that the
fusion center has access to the physical locations of the sensors
and can correctly associate messages with their points of origin.
This may be justifiable by possible models for the underlying
wireless transmission where triangulation of sensors is inherently
performed. The problem setup is illustrated in
Figure~\ref{fig:mainFig}.

The available literature on distributed field estimation which
simultaneously treats binary--{\em sensing}, random sensor {\em
deployment}, and unknown observation noise distribution is limited.
The early works in \cite{Masry-IT1981-RASFS,MasryC-IT1981-BPCNT}
consider the problem of reconstructing a signal from
binary--quantized samples acquired with random thresholds, but do
not consider arbitrary additive noise with an unknown distribution
and only consider fixed deterministic sampling locations
(deployment). The work in \cite{Luo-IT2005-UDEBCSN} is limited to
the estimation of a {\em constant} field and does not explicitly
address sampling precision (sensing) constraints. A recent work
\cite{WangMZIS-Allerton06-UDEoNFOb} provides pointwise MSE decay
rates in terms of the local and global modulus of field continuity
by building upon the techniques in
\cite{Masry-IT1981-RASFS,MasryC-IT1981-BPCNT,Luo-IT2005-UDEBCSN}.
However, \cite{WangMZIS-Allerton06-UDEoNFOb} does not consider
random sensor deployment and requires local field continuity for
pointwise MSE convergence.  The present work incorporates random
sensor deployment, binary--sensing, and unknown noise distribution
while studying almost sure and integrated MSE convergence of the
field estimate.  The integrated MSE convergence for the
bounded--variation, Sobolev--differentiable, and finite--dimensional
function classes are explored in detail.  Our results expose the
effects of field ``smoothness'', deployment randomness, and
observation/sensing noise on the integrated MSE scaling behavior.

For field estimation approaches which are not constrained by finite
sensing precision and sensing unreliability, such as those involving
``uncoded'' analog joint sampling--transmission, there is a growing
body of literature now available (e.g., see
\cite{GastparV-2003-SCCSN, GastparRV-2003-TCNCLSCCR,
NowakMW-2004-EIFWSN, LiuES-2005-OPFESN, BajwaSN-2005-MSCCFEWSN,
LiuU-2006-ODPTGSN, Gastpar-ITA07-UTEOSGSN} and references therein).
Related to the distributed field reconstruction problem is the
so--called CEO problem studied in the Information Theory community
in which the distortion is averaged over multiple field snapshots
over time (e.g., see
\cite{BergerZV-IT1996-TCP,ViswanathanB-IT1997-QGCP,PrabhakaranTR-ISIT04-RQGCP}
and references therein). There is also a significant body of work on
oversampled A--D conversion (e.g., see
\cite{Cvetkovic-IT2003-RPREUANQ} and references therein), which is
loosely related to the results of the present work concerning
finite--dimensional fields.  However, these are different problem
formulations and are not the focus of the present work.

The rest of this paper is organized as follows. The problem
formulation with detailed modeling assumptions are presented in
Section~\ref{sec:problemSetup}. The core technical results are then
summarized and discussed in Section~\ref{sec:mainResults}. The core
results are then used to derived explicit expressions of the decay
rate of the integrated MSE for three rich function classes in
Section~\ref{sec:MSEDecayRates}. The proofs of all the core
technical results are presented in Section~\ref{sec:proofs} and
concluding remarks are made in Section~\ref{sec:concl}.

\section{Problem Formulation} \label{sec:problemSetup}
\textbf{Field Model:} We model the field as a real--valued, bounded,
deterministic function $f : \mathcal{D} \rightarrow [-a,+a]$
belonging to a non--parametric function class\footnote{The number of
parameters that specify a non--parametric function class is not
fixed a priori and is possibly infinite.} $\mathcal{F}$, that is, $f
\in \mathcal{F}$, where $\mathcal{F}$ is a set of measurable
functions mapping $\mathcal{D}$ to $[-a,+a]$. The domain of the
field $\mathcal{D}$ is assumed to be a compact subset of $\rR^d$,
the $d$--dimensional Euclidean space. The objective is to
reconstruct this function with high fidelity from binary--quantized
noisy observations collected by a network of
non--cooperative\footnote{The sensors do not exchange information or
otherwise collaborate at the time of or before taking measurements.}
sensors that are randomly deployed over the domain $\mathcal{D}$.

\textbf{Random Sensor Deployment:} We assume that the $n$ sensors
are independently and identically randomly deployed over the domain
$\mathcal{D}$ according to a known distribution $p_X$. If $X_i \in
\mathcal{D}$ denotes the location of the $i^{\textrm{th}}$ sensor
for $i \in \{1, \ldots, n\}$, then $X_i \sim \mbox{iid }p_X$
captures the lack of control in sensor deployment. We assume that
the support of $p_X$ is $\mathcal{D}$ and that $p_X$ is a
non--singular distribution\footnote{A random variable with a
non--singular distribution takes values in a subset of $\mathcal{D}$
with Lebesgue measure $0$ with probability $0$.}.

\textbf{Additive Noise:} Each sensor takes a sample of the field
under additive noise. The noisy samples are given by $Y_i = f(X_i) +
Z_i$, for $i \in \{1, \ldots, n\}$, where the noise variables $Z_i
\sim \mbox{iid } p_Z$ and are independent of the sensor locations.
We assume that each $Z_i$ is zero--mean and is bounded in amplitude
by a constant $b > 0$, that is, the support of $p_Z$ is contained in
$[-b,+b]$. However, besides these assumed conditions, the
distribution $p_Z$ is {\em unknown} to either the sensors or the
fusion center, and the results and methods of this paper hold for
{\em arbitrary} noise distributions satisfying these conditions. We
let $\mathcal{P}_Z$ denote the set of all noise distributions
satisfying these assumptions. Note that since both the field and the
noise are bounded, the noisy samples are bounded: $|Y_i| \leq c := a
+ b$. We assume that the value of $c$ is known. The values of $a$
and $b$ can remain unknown to the sensors and the fusion center.

\textbf{Unreliable, Binary Quantization:} We assume that in the
sensor hardware frontend, the noisy sample is quantized by an
unreliable, low--precision analog--to--digital converter.
Specifically, we consider one--bit (binary), dithered, scalar
quantization implemented as a comparison to a random threshold that
is uniformly distributed over the sample dynamic range $[-c,+c]$.
The binary--quantized observations are given by
\begin{eqnarray*}
B_i &=& \mathrm{sgn}(Y_i - T_i) \quad \quad \quad \mbox{for } i \in \{1, \ldots, n\} \\
&:=& \begin{cases} +1 & Y_i > T_i, \\
-1, & Y_i \leq T_i,
\end{cases}
= \begin{cases} +1 & f(X_i) + Z_i > T_i, \\
-1, & f(X_i) + Z_i \leq T_i,
\end{cases}
\end{eqnarray*}
where $T_i \sim \mbox{iid Unif}[-c,+c]$ are the uniform random
thresholds. The thresholds are independent of the sensor locations
and the noise. The value of $B_i$ is finally the observation that
sensor $i$ has access to.

\textbf{Transmission:} We abstract away communication channel issues
and assume that the underlying communication network of the sensors
is able to handle the modest payload of transmitting one bit per
sensor to the fusion center. We also assume that the fusion center
has access to the physical locations of the sensors and can
correctly associate messages with their points of origin. Thus, we
assume that through this abstracted communications network, the
sensor location and quantized observation pairs
$\{(X_i,B_i)\}_{i=1}^n$ are reliably made available to the fusion
center. The reconstruction of multiple time snapshots of the field
with time--sharing sensors, vanishing per--sensor rate and sensor
location ``overheads'' can also be accommodated within the framework
of this work as in \cite{WangMZIS-Allerton06-UDEoNFOb} but is
omitted for clarity.

\textbf{Reconstruction and Distortion Criterion:} Given
$\{(X_i,B_i)\}_{i=1}^n$, the fusion center constructs the field
estimate $\hat{f}_{X_1, \ldots, X_n, B_1, \ldots, B_n} : \mathcal{D}
\rightarrow \cC$. For notational convenience, the explicit
dependence on $\{(X_i,B_i)\}_{i=1}^n$ will be suppressed and the
estimator will simply be denoted by $\hat{f}_n$. The performance
criterion is the integrated MSE given by
\[
D(f,\hat{f}_n) := \eE \left[\|f - \hat{f}_n\|^2 \right] = \eE \left[
\int_{\mathcal{D}} |f(x) - \hat{f}_n(x)|^2 dx \right],
\]
where the expectation is taken with respect to the random noise,
thresholds, and the sensor locations. The objective is to design an
estimator $\hat{f}_n$ that minimizes the integrated MSE $D$. The
problem setup is shown in \figref{fig:mainFig}.

\textbf{Minimax Integrated MSE:} For a given field subclass
$\mathcal{F}_{\mathrm{sub}} \subset \mathcal{F}$, of interest are
the corresponding upper, lower, and minimax rates of convergence of
the integrated MSE. A positive sequence $\gamma_n$ is an upper rate
of convergence if there exists a constant $C < \infty$ and an
estimator $\hat{f}_n^*$ such that
\[
\limsup_{n \longrightarrow \infty} \sup_{p_Z \in \mathcal{P}_Z}
\sup_{f \in \mathcal{F}_{\mathrm{sub}}} \gamma_n^{-1}
D(f,\hat{f}_n^*) \leq C.
\]
A positive sequence $\gamma_n$ is a lower rate of convergence if
there exists a constant $C > 0$ such that
\[
\liminf_{n \longrightarrow \infty} \inf_{\hat{f}_n} \sup_{p_Z \in
\mathcal{P}_Z} \sup_{f \in \mathcal{F}_{\mathrm{sub}}} \gamma_n^{-1}
D(f,\hat{f}_n) \geq C,
\]
where the $\inf_{\hat{f}_n}$ is the infimum over all field
estimators. The upper rate represents the asymptotic worst--case
performance achieved by a given estimator. The lower rate represents
a fundamental limit on the asymptotic performance of any estimator.
A positive sequence $\gamma_n$ that is both a lower rate and an
upper rate of convergence is called the minimax rate of convergence
and the corresponding estimator $\hat{f}_n^*$ that achieves the
upper rate is called a minimax order optimal estimator.

Note that showing $D(f,\hat{f}_n^*) = O(\gamma_n)$ for all $f \in
\mathcal{F}_{\mathrm{sub}}$ and $p_Z \in \mathcal{P}_Z$ for a
particular estimator $\hat{f}_n^*$ is equivalent to showing that
$\hat{f}_n^*$ achieves $\gamma_n$ as an upper rate of convergence of
the integrated MSE. If it can be further shown that $D(f,\hat{f}_n)
= \Omega(\gamma_n)$ for a particular $f \in
\mathcal{F}_{\mathrm{sub}}$, a particular $p_Z \in \mathcal{P}_Z$,
and for all estimators $\hat{f}_n$, then $\gamma_n$ is the minimax
rate of convergence of the integrated MSE.

\section{Main Results} \label{sec:mainResults}
In this section, we describe our proposed field estimator and
analyze its performance. We show that under suitable technical
conditions, the field estimate is asymptotically integrated MSE
consistent, that is, as $n \longrightarrow \infty$, $\eE[\|f -
\hat{f}_n\|^2] \longrightarrow 0$. We also show that under suitable
technical conditions, the field estimate is asymptotically almost
sure consistent, that is, as $n \longrightarrow \infty$, almost
surely $\hat{f}_n \longrightarrow f$ pointwise almost everywhere on
$\mathcal{D}$. We also provide an upper bound to the integrated MSE
which is used in \secref{sec:MSEDecayRates} to derive achievable
integrated MSE decay rates for specific function classes. The proofs
of all theorems are presented in \secref{sec:proofs}.

Let $\mathcal{F}$ denote the set of all bounded, measurable
functions $f : \mathcal{D} \rightarrow [-a,+a]$. Note that
$\mathcal{F} \subseteq \mathbf{L}^2(\mathcal{D})$. Let $\mathcal{B}
= \{\phi_j\}_{j=0}^{\infty}$, with $\phi_j : \mathcal{D} \rightarrow
\cC$, denote an indexed orthonormal (Schauder) basis (e.g. Fourier,
wavelet, etc.) of $\mathbf{L}^2(\mathcal{D})$. Any $f \in
\mathcal{F}$ can be decomposed as
\begin{equation} \label{eqn:decomposition}
f \stackrel{\mathbf{L}^2}{=} \sum_{j = 0}^{\infty} \langle f,\phi_j
\rangle \phi_j =: \sum_{j = 0}^{\infty} \alpha_j \phi_j,
\end{equation}
where $\alpha_j := \langle f,\phi_j \rangle$ denotes the
coefficients (projections onto the basis functions) of the
expansion. The $m$--term approximation of $f$ with respect to an
orthonormal basis $\mathcal{B} = \{\phi_j\}_{j=0}^{\infty}$ is given
by
\begin{equation} \label{eqn:mTermApprox}
f_m := \sum_{j = 0}^{m-1} \langle f,\phi_j \rangle \phi_j.
\end{equation}
The corresponding $m$--term approximation error is given by
\begin{equation} \label{eqn:mTermError}
\varepsilon[f,m,\mathcal{B}] := \|f-f_m\|^2 = \sum_{j = m}^{\infty}
|\langle f,\phi_j \rangle|^2 = \sum_{j=m}^{\infty} |\alpha_j|^2,
\end{equation}
which is a non--negative, non--increasing sequence of $m$ that
converges to zero for all $f \in \mathcal{F}$
\cite[Chapter~9]{Mallat-AP98-WToSP}.

\subsection{Proposed estimator} Our proposed estimator first
estimates the first $m$ coefficients $\{ \alpha_j \}_{j=0}^{m-1}$ of
\eqnref{eqn:decomposition} with respect to a given orthonormal basis
$\mathcal{B}$, according to
\begin{equation} \label{eqn:coefEst}
\hat{\alpha}_j := \frac{c}{n} \sum_{i=1}^n
\frac{\phi_j^*(X_i)}{p_X(X_i)} B_i,
\end{equation}
for $j \in \{0, \ldots, m-1\}$. A general tunable field estimate is
given by the $m$--term approximation,
\begin{equation} \label{eqn:finiteMEstFunc}
\hat{f}_{n,m} := \sum_{j = 0}^{m-1} \hat{\alpha}_j \phi_j,
\end{equation}
where $m$ is the tunable design parameter which can be chosen to
depend on $n$ to optimize the rate of decay of the integrated MSE
for specific function classes. The final field estimate is given by
specifying $m$ as a function of $n$,
\begin{equation} \label{eqn:EstFunc}
\hat{f}_{n} := \sum_{j = 0}^{m(n)-1} \hat{\alpha}_j \phi_j.
\end{equation}
The specification of $m(n)$ for specific function classes is
discussed in \secref{sec:MSEDecayRates}. The dependence of $m$ on
$n$ needs to satisfy certain conditions to ensure that the estimate
is asymptotically consistent. These conditions are described in
\secref{sec:perfResults} and \secref{sec:ASConvResults}.

\subsection{Integrated MSE upper bounds and convergence results} \label{sec:perfResults}
The following theorem, whose proof appears in
\secref{sec:MainProof}, upper bounds the integrated MSE as the sum
of two terms. The first term is due to the variance of the
coefficient estimates. The second term is due to the bias caused by
the finite--term series approximation.

{\bf
\begin{theorem} \label{thm:MSEUpperBound}
{\it \textbf{(Integrated MSE Upper Bound)} Let $\mathcal{F}$,
$\mathcal{P}_Z$, and $p_X$ be as given in \secref{sec:problemSetup}.
Let $\hat{f}_{n,m}$ be given by \eqnref{eqn:coefEst} and
\eqnref{eqn:finiteMEstFunc}, where $\mathcal{B} =
\{\phi_j\}_{j=0}^{\infty}$ is any orthonormal Schauder basis of
$\mathbf{L}^2(\mathcal{D})$. Then, $\forall f \in \mathcal{F}$ and
$\forall p_Z \in \mathcal{P}_Z$, the integrated MSE is upper bounded
by
\begin{equation} \label{eqn:MSEUpperBound}
D = \eE \left[ \|f - \hat{f}_{n,m}\|^2 \right] \leq \frac{c^2}{n}
\sum_{j=0}^{m-1} \int_{\mathcal{D}} \frac{|\phi_j(x)|^2}{p_X(x)} dx
+ \varepsilon[f,m,\mathcal{B}],
\end{equation}
where $\varepsilon[f,m,\mathcal{B}]$, given by
\eqnref{eqn:mTermError}, is a non--negative, non--increasing
sequence that converges to $0$ as $m \longrightarrow \infty$.}
\end{theorem}
}

In light of \thmref{thm:MSEUpperBound} we now examine conditions on
$m(n)$, $\mathcal{B}$, and $p_X$ which ensure that the estimator is
asymptotically consistent in the integrated MSE sense, that is $D
\longrightarrow 0$ as $n \longrightarrow \infty$. The following
corollary specifies conditions that immediately ensures that the
integrated MSE converges to $0$.

{\bf
\begin{corollary} \label{thm:MSEGoesToZero}
{\it \textbf{(Integrated MSE Convergence of the Field Estimate)}
Under the same setup of \thmref{thm:MSEUpperBound}, if $m(n)$,
$\mathcal{B} = \{\phi_j\}_{j=0}^{\infty}$, and $p_X$ satisfy
\begin{eqnarray} \label{eqn:MSEConvCond1}
&&m(n) \longrightarrow \infty,\quad \mbox{as }
n \longrightarrow \infty, \\
&&\frac{1}{n}\sum_{j=0}^{m(n)-1} \int_{\mathcal{D}}
\frac{|\phi_j(x)|^2}{p_X(x)} dx \longrightarrow 0, \quad \mbox{as }
n \longrightarrow \infty, \label{eqn:MSEConvCond2}
\end{eqnarray}
then the estimate converges in the integrated MSE sense to the
field, that is,
\[
\forall f \in \mathcal{F} \mbox{ and } \forall p_Z \in \mathcal{P}_Z, \quad
D \longrightarrow 0, \quad \mbox{as } n \longrightarrow \infty.
\] }
\end{corollary}
}

Condition \eqnref{eqn:MSEConvCond1} is sufficient (and often
necessary) to ensure that $\varepsilon[f,m,\mathcal{B}]$ converges
to $0$. Condition \eqnref{eqn:MSEConvCond2} is equivalent to the
first term of the integrated MSE upper bound, given in
\eqnref{eqn:MSEUpperBound}, converging to $0$. For some deployment
distributions $p_X$, condition \eqnref{eqn:MSEConvCond2} may not be
attainable for many orthonormal bases. For example, let the domain
$\mathcal{D} = [0,1]$ with the deployment distribution $p_X(x) = 2x$
over $[0,1]$. Then for any orthonormal basis in which $\phi_0(x) =
1$ over $[0,1]$, e.g., Fourier, Harr wavelets, Legendre polynomials,
etc., the first term of the summation in \eqnref{eqn:MSEConvCond2}
is given by
\[
\int_{\mathcal{D}} \frac{|\phi_0(x)|^2}{p_X(x)} dx = \int_0^1
\frac{1}{2x} dx = \infty.
\]
Thus integrated MSE upper bound becomes useless. This implies that in
general the deployment distributions and orthonormal bases have to be
appropriately \emph{matched} as a design consideration in order to
satisfy condition \eqnref{eqn:MSEConvCond2}. However, condition
\eqnref{eqn:MSEConvCond2} is ensured for any orthonormal basis if the
deployment distribution $p_X$ has a strictly positive infimum over
$\mathcal{D}$, that is,
\begin{equation} \label{eqn:deployCond}
\inf_{x \in \mathcal{D}} p_X(x) = \nu > 0.
\end{equation}
Sensor deployment distributions over compact domains which are
useful for high-resolution field reconstruction would satisfy such a
condition.  Given \eqnref{eqn:deployCond}, we have that
\begin{eqnarray*}
\sum_{j=0}^{m-1} \int_{\mathcal{D}} \frac{|\phi_j(x)|^2}{p_X(x)} dx
&\leq& \sum_{j=0}^{m-1} \int_{\mathcal{D}} \frac{|\phi_j(x)|^2}{\nu}
dx \\
&=& \sum_{j=0}^{m-1} \frac{1}{\nu} \|\phi_j\|^2 = \frac{m}{\nu},
\end{eqnarray*}
and $ \forall f \in \mathcal{F}$ and $\forall p_Z \in \mathcal{P}_Z$,
the corresponding integrated MSE upper bound becomes
\begin{equation} \label{eqn:MSEBoundReduced}
D = \eE \left[ \|f - \hat{f}_{n,m}\|^2 \right] \leq
\frac{c^2m}{n\nu} + \varepsilon[f,m,\mathcal{B}].
\end{equation}
The upper bound in \eqnref{eqn:MSEBoundReduced} converges to $0$ as
$n \longrightarrow \infty$ when $m(n)$ satisfies the following two
conditions
\begin{eqnarray*}
&& m(n) \longrightarrow \infty, \quad \mbox{as } n \longrightarrow \infty, \\
&& \frac{m(n)}{n} \longrightarrow 0, \quad \mbox{as } n
\longrightarrow \infty.
\end{eqnarray*}

\subsection{Almost sure convergence results} \label{sec:ASConvResults}
In this subsection we establish sufficient conditions for the field
estimate to be asymptotically almost sure consistent, that is, as $n
\longrightarrow \infty$, almost surely $\hat{f}_n \longrightarrow f$
pointwise almost everywhere on $\mathcal{D}$. First, we establish a
key theorem that gives sufficient conditions for the convergence of
the pointwise errors of the estimate with respect to the truncated
approximation of the field. The proof of this theorem appears in
\secref{sec:AsErrorProof}.

{\bf
\begin{theorem} \label{thm:ErrorASConv}
{\it \textbf{(Almost Sure Convergence of Estimate Errors)} Let $p_X$
be the deployment distribution described in
\secref{sec:problemSetup} satisfying \eqnref{eqn:deployCond},
$\mathcal{B} = \{\phi_j\}_{j=0}^{\infty}$ be an orthonormal Schauder
basis of $\mathbf{L}^2(\mathcal{D})$, $\hat{f}_{n,m}$ be the field
estimate given by \eqnref{eqn:coefEst} and
\eqnref{eqn:finiteMEstFunc}, and $f_{m}$ be the $m$--term
approximation to the field given by \eqnref{eqn:mTermApprox}. Let
\[
S_{n,m}(x) := \hat{f}_{n,m}(x) - f_{m}(x),
\]
for all $x \in \mathcal{D}$. If there exists a non--negative,
increasing sequence of real numbers $\{\Lambda_m\}_{m=1}^{\infty}$,
and a non--negative, increasing sequence of positive integers
$\{m(n)\}_{n=1}^{\infty}$ which satisfy the following three
conditions
\begin{align}
\forall x,y \stackrel{\mathrm{a.e.}}{\in}
\mathcal{D}, &\left|\sum_{j=0}^{m-1} \phi_j(x) \phi_j^*(y)\frac{1}{p_X(y)}\right|
\leq C_1 \Lambda_m,
\label{eqn:ASConvCond1} \\
\forall f \in \mathcal{F}, \forall x
\stackrel{\mathrm{a.e.}}{\in} \mathcal{D},
&\left|\sum_{j=0}^{m-1} \langle f,\phi_j \rangle \phi_j(x) \right|
\leq C_2 \Lambda_m,   \label{eqn:ASConvCond2} \\
\forall \epsilon > 0,
&\sum_{n=1}^{\infty} \exp\left( \frac{- \epsilon^2
n}{\Lambda_{m(n)}^2} \right) < \infty,
\label{eqn:ASConvCond3}
\end{align}
where $C_1,C_2 > 0$ are some constants, then $\forall f \in
\mathcal{F}$ and $\forall x \in \mathcal{D}$ except on a set of
Lebesgue measure zero, as $n \longrightarrow \infty$, almost surely,
\[
S_n(x) := S_{n,m(n)} \longrightarrow 0.
\]}
\end{theorem}
}

Conditions \eqnref{eqn:ASConvCond1} and \eqnref{eqn:ASConvCond2}
impose constraints on the basis functions $\mathcal{B} =
\{\phi_j\}_{j=0}^{\infty}$ and the deployment distribution $p_X$.
Condition \eqnref{eqn:ASConvCond3} implies that as $n
\longrightarrow \infty$, $\Lambda_{m(n)}^2/n \longrightarrow 0$.
This places a constraint on how fast $m(n)$ can go to infinity. In
particular it requires that in relation to $\Lambda_m$, $m(n)$ not
grow too fast with $n$.

We now examine some special choices of
$\{\Lambda_m\}_{m=1}^{\infty}$ for which conditions
\eqnref{eqn:ASConvCond1} and \eqnref{eqn:ASConvCond2} will hold. For
$m \in \{1, 2, \ldots\}$, define auxiliary functions:
\begin{eqnarray}
&& g_m(x,y) := \frac{1}{\Lambda_m} \sum_{j=0}^{m-1} \phi_j(x)
\phi_j^*(y)\frac{1}{p_X(y)}, \label{eqn:gAuxFcn} \\
&& h_m(x) := \frac{1}{\Lambda_m} \sum_{j=0}^{m-1} \langle f,\phi_j
\rangle \phi_j(x), \label{eqn:hAuxFcn}
\end{eqnarray}
for $x, y \in \mathcal{D}$. The following proposition, whose proof
appears in \secref{sec:PropProof}, gives two sets of conditions on
$\{\Lambda_m\}_{m=1}^{\infty}$, $\mathcal{B} =
\{\phi_j\}_{j=0}^{\infty}$, and $p_X$, for which conditions
\eqnref{eqn:ASConvCond1} and \eqnref{eqn:ASConvCond2} will hold.

{\bf
\begin{proposition} \label{prp:asAuxFcnsPrp}
{\it Let $\{\Lambda_m\}_{m=1}^{\infty}$ be as in
\thmref{thm:ErrorASConv} and $g_m$, $h_m$ be given by
\eqnref{eqn:gAuxFcn} and \eqnref{eqn:hAuxFcn} respectively.
\begin{itemize}
\item[(i)] If
\begin{equation} \label{eqn:ASBoundCond}
\frac{m}{\Lambda_m^2} \longrightarrow 0, \quad \mbox{as } m
\longrightarrow \infty,
\end{equation}
and for $x, y \in \mathcal{D}$ almost everywhere,
the limits
\begin{eqnarray}
&& g_{\infty}(x,y) := \lim_{m \longrightarrow \infty}
g_m(x,y), \label{eqn:limsAuxFuns1} \\
&& h_{\infty}(x) := \lim_{m \longrightarrow \infty} h_m(x)
\label{eqn:limsAuxFuns}
\end{eqnarray}
exist, then the limits are zero almost everywhere and the conditions
\eqnref{eqn:ASConvCond1} and \eqnref{eqn:ASConvCond2} are satisfied
for some constants $C_1, C_2 > 0$.
\item[(ii)] If the basis functions are uniformly amplitude bounded,
that is, $\forall j \in \{0, 1, \ldots\}$ and $ \forall x \in
\mathcal{D}$,
\[
|\phi_j(x)| \leq \beta < \infty,
\]
then conditions \eqnref{eqn:ASConvCond1} and
\eqnref{eqn:ASConvCond2} are satisfied for $\Lambda_m = m$ with
constants $C_1 = \beta^2/\nu$ and $C_2 = a \beta
\sqrt{\mathrm{vol}(\mathcal{D})}$.
\end{itemize} }
\end{proposition}
}

Part (i) of \prpref{prp:asAuxFcnsPrp} shows that if the limits of the
auxiliary functions \eqnref{eqn:gAuxFcn} and \eqnref{eqn:hAuxFcn} as
$m \longrightarrow \infty$ exist, then for any $\Lambda_m$ such that
\eqnref{eqn:ASBoundCond} is satisfied, e.g., $\Lambda_m =
m^{\gamma/2}$, for any $\gamma > 1$, conditions
\eqnref{eqn:ASConvCond1} and \eqnref{eqn:ASConvCond2} are satisfied
for some constants. Part (ii) of \prpref{prp:asAuxFcnsPrp} shows that
if the basis functions are uniformly bounded as, for example, in the
orthonormal Fourier and Legendre bases, then conditions
\eqnref{eqn:ASConvCond1} and \eqnref{eqn:ASConvCond2} are satisfied
for $\Lambda_m = m$ and given constants.

We now examine conditions on $\{\Lambda_m\}_{m=1}^{\infty}$ under
which \eqnref{eqn:ASConvCond3} will be satisfied. According to
Ermakoff's series convergence test \cite{Knopp-Dover90-TaAoIS}, if
for some non--negative, non--increasing, real function $q(t)$, $t
\geq 1$,
\[
\lim_{t \longrightarrow \infty} \frac{e^t q(e^t)}{q(t)} < 1,
\]
where $e$ is the base of the natural logarithm, then
\[
\sum_{n=1}^\infty q(n) < \infty.
\]
Let $q(t) = \exp \big(\frac{- \epsilon^2 t}{t^\psi}\big)$, $t \geq
1$, where $\psi \in (0,1)$ and $\epsilon > 0$. Then
\[
\frac{e^t q(e^t)}{q(t)} = \frac{e^t \exp(\frac{-\epsilon^2
e^t}{e^{t\psi}})}{\exp(\frac{-\epsilon^2 t}{t^\psi})} = \exp\left(
t- \epsilon^2 e^{t - t\psi} - \epsilon^2 t^{1-\psi} \right)
\longrightarrow 0,
\]
as $t \longrightarrow \infty$.
By Ermakoff's test, for all $\psi \in (0,1)$ and all $\epsilon > 0$,
\[
\sum_{n=1}^{\infty} \exp\left( \frac{- \epsilon^2 n}{n^{\psi}}
\right) < \infty.
\]
Thus condition \eqnref{eqn:ASConvCond3} will be satisfied if
$\Lambda_{m(n)}^2 = n^\psi$ for any $\psi \in (0,1)$.

Combining the above result with \prpref{prp:asAuxFcnsPrp} yields
possible forms of the design parameters $\{m(n)\}_{n=1}^{\infty}$
and $\{\Lambda_m\}_{m=1}^{\infty}$ such that the conditions for
almost sure convergence \eqnref{eqn:ASConvCond1},
\eqnref{eqn:ASConvCond2}, and \eqnref{eqn:ASConvCond3} are all
simultaneously satisfied. Choosing $m(n) = \Theta(n^{\psi})$, where
$\psi \in (0,1)$, and $\Lambda_m = m^{\gamma/2}$, for some $\gamma
\in (1,1/\psi)$, yields $\Lambda_{m(n)}^2 = n^{\psi^\prime}$, where
$\psi^\prime = \gamma \psi \in (\psi,1)$, which satisfies
\eqnref{eqn:ASConvCond3} and \eqnref{eqn:ASBoundCond}
simultaneously. With these choices, \prpref{prp:asAuxFcnsPrp} shows
that conditions \eqnref{eqn:ASConvCond1} and
\eqnref{eqn:ASConvCond2} will be satisfied as well if the limits
\eqnref{eqn:limsAuxFuns1} and \eqnref{eqn:limsAuxFuns} of the
auxiliary functions \eqnref{eqn:gAuxFcn} and \eqnref{eqn:hAuxFcn}
respectively can be assumed to exist. Thus, for any $m(n)$ of the
form $m(n) = \Theta(n^{\psi})$, where $\psi \in (0,1)$, we can
choose $\{\Lambda_m\}_{m=1}^{\infty}$ such that conditions
\eqnref{eqn:ASConvCond1}, \eqnref{eqn:ASConvCond2}, and
\eqnref{eqn:ASConvCond3} are simultaneously satisfied, if the limits
\eqnref{eqn:limsAuxFuns1} and \eqnref{eqn:limsAuxFuns} exist.

Due to the properties of an orthonormal basis, as $m \longrightarrow
\infty$, the $m$--term approximation, $f_m$ given by
\eqnref{eqn:mTermApprox}, converges in $\mathbf{L}^2$--norm to $f$
for any $f \in \mathcal{F}$. Although, it is not guaranteed that for
general orthonormal bases $f_m$ will converge pointwise almost
everywhere to a specific function. However, if $f_m$ does converge
almost everywhere to some $f_\infty$, then $f_\infty$ must be equal
to $f$ almost everywhere. This can be seen by writing
\begin{eqnarray*}
0 &\leq& \int_{\mathcal{D}} |f(x) - f_{\infty}(x)|^2 dx \\
&=& \int_{\mathcal{D}} \liminf_{m \longrightarrow \infty} |f(x) -
f_{m}(x)|^2 dx \\
&\leq& \liminf_{m \longrightarrow \infty} \int_{\mathcal{D}} |f(x) -
f_{m}(x)|^2 dx = 0,
\end{eqnarray*}
where the inequality follows due to Fatou's lemma
\cite{Rudin-McGraw74-RaCA}. Thus $\int_{\mathcal{D}} |f(x) -
f_\infty(x)|^2 dx = 0$, so $|f(x) - f_\infty(x)| = 0$ for $x \in
\mathcal{D}$ almost everywhere. For example, it is well known that
for any $f \in \mathcal{F} \subset \mathbf{L}^2([0,1])$ the
$m$--term Fourier series approximation, $f_{m}$ converges to $f$
almost everywhere \cite{Carleson-ActaMath1966-CaGPSFS}.

{\bf
\begin{corollary}\label{thm:EstASConv}
{\it \textbf{(Almost Sure Convergence of the Field Estimate)} Within
the context of \thmref{thm:ErrorASConv}, if conditions
\eqnref{eqn:ASConvCond1}, \eqnref{eqn:ASConvCond2}, and
\eqnref{eqn:ASConvCond3} hold and if as $m \longrightarrow \infty$,
the $m$--term approximation $f_{m}$ converges almost everywhere to
some function $f_\infty$, then $\forall x
\stackrel{\mathrm{a.e.}}{\in} \mathcal{D}$, the pointwise error of the
field estimate satisfies
\begin{eqnarray*}
|\hat{f}_{n}(x) - f(x)| &\leq& |\hat{f}_{n}(x) - f_{m(n)}(x)| +
|f_{m(n)}(x) - f(x)| \\
&=& |S_n(x)| + |f_{m(n)}(x) - f(x)| \\
&&\xrightarrow{\mathrm{a.s.}} 0, \quad \mbox{as } n \longrightarrow
\infty.
\end{eqnarray*}
Thus, for $x \in \mathcal{D}$ almost everywhere, as $n \longrightarrow
\infty$, almost surely $\hat{f}_{n}(x) \longrightarrow f(x)$.}
\end{corollary}
}

\section{Achievable Integrated MSE Decay Rates} \label{sec:MSEDecayRates}
In this section, we use the integrated MSE upper bound
\eqnref{eqn:MSEUpperBound} derive explicit expressions for the
achievable upper rates of convergence of the integrated MSE for
three specific function classes, namely, finite--dimensional
$\mathcal{F}_{\mathcal{B}_k}$, bounded--variation
$\mathcal{F}_{BV}$, and $s$--Sobolev differentiable $\mathcal{F}_s$.
Throughout this section, we assume that \eqnref{eqn:deployCond}
holds.

The general approach for deriving such rates of convergence for
functions living in a function class $\mathcal{F}_{\mathrm{sub}}
\subseteq \mathcal{F}$ is select an appropriate basis $\mathcal{B} =
\{\phi_j\}_{j=0}^{\infty}$ in which the $m$--term approximation
error given by $\varepsilon[f,m,\mathcal{B}]$ in
\eqnref{eqn:mTermError} can be upper bounded by an explicit function
of $m$ for all $f \in \mathcal{F}_{\mathrm{sub}}$. Then $m$ in
\eqnref{eqn:MSEUpperBound} can be chosen to depend on $n$ to
optimize the convergence rate. Thus given the appropriate function
approximation theoretic results that upper bound
$\varepsilon[f,m,\mathcal{B}]$, this approach establishes achievable
upper rates of convergence of the integrated MSE for the
corresponding function class.

\subsection{Functions in a finite--dimensional subspace of $\mathcal{F}$}
The first function class represents the scenario where the fusion
center has an exact prior knowledge of the finite--dimensional space
in which the function lives. Let $\mathcal{F}_{\mathcal{B}_k}$
denote the subset of $\mathcal{F}$ that is composed of functions
that are linear combinations of a given set of $k$ orthonormal
functions $\mathcal{B}_k = \{\phi_j\}_{j=0}^{k-1}$. Note that for
any $f \in \mathcal{F}_{\mathcal{B}_k}$, $f = \sum_{j = 0}^{k-1}
\langle f,\phi_j \rangle \phi_j$. Thus the function approximation at
the truncation point $m = k$ is exact, that is, $f_m = f$ for $m =k$
so that,
\begin{equation} \label{eqn:varepsFiniteD}
\forall f \in \mathcal{F}_{\mathcal{B}_k}, \quad
\varepsilon[f,k,\mathcal{B}_k] = 0.
\end{equation}
Combining \eqnref{eqn:MSEUpperBound} with \eqnref{eqn:varepsFiniteD}
yields the following corollary.

{\bf
\begin{corollary} \label{thm:MSERateFiniteD}
{\it \textbf{(Decay rate of integrated MSE for
$\mathcal{F}_{\mathcal{B}_k}$)} Let $\mathcal{B}_k$ and
$\mathcal{F}_{\mathcal{B}_k}$ be as given above and $\mathcal{P}_Z$
and $p_X$ be as given in \secref{sec:problemSetup}. Let
$\hat{f}_{n,m}$ be given by \eqnref{eqn:coefEst} and
\eqnref{eqn:finiteMEstFunc} with $\mathcal{B}_k$ as the basis. If
$p_X$ satisfies \eqnref{eqn:deployCond}, then $\forall f \in
\mathcal{F}_{\mathcal{B}_k}$ and $\forall p_Z \in \mathcal{P}_Z$, the
integrated MSE of $\hat{f}_{n,m}$ with the truncation point $m$ set to
$k$ is upper bounded as follows
\[
D = \eE \left[ \|f - \hat{f}_{n,m}\|^2 \right] \leq \frac{c^2 k}{n
\nu} = O \left(\frac{1}{n} \right).
\]
Therefore, $\forall f \in \mathcal{F}_{\mathcal{B}_k}$ and $\forall
p_Z \in \mathcal{P}_Z$, an achievable upper rate of convergence of the
integrated MSE for fields in a finite--dimensional subspace is given
by
\[
D = \eE \left[ \|f - \hat{f}_{n,m}\|^2 \right] = O \left(\frac{1}{n}
\right).
\]}
\end{corollary}
}

It should be noted that for this function class, the field estimation
problem for integrated MSE is equivalent to a finite--dimensional
parameter estimation problem with conditionally independent noisy
observations. Under the choice of an appropriate, well--behaved noise
distribution\footnote{A noise distribution is chosen such that the
observation model satisfies the Cram\'{e}r--Rao regularity conditions
\cite{Kay-1993-FSSPET}.} $p_Z \in \mathcal{P}_Z$, the Cram\'{e}r--Rao
lower bound for the integrated MSE decay rate for finite--dimensional
parameter estimation from iid noisy observations asymptotically
behaves as $D = \Omega(1/n)$ for all asymptotically integrated MSE
consistent estimators \cite{Kay-1993-FSSPET}. Hence the estimator is
minimax order optimal for $\mathcal{F}_{\mathcal{B}_k}$ and achieves
the minimax rate of convergence $\gamma_n = (1/n)$.

\subsection{Functions of bounded--variation on Domain $\mathcal{D} = [0,1]$}
Let $\mathcal{F}_{BV}$ denote the subset of $\mathcal{F}$ which is
composed of functions on $\mathcal{D} = [0,1]$ of
bounded--variation. Formally,
\[
\mathcal{F}_{BV} := \left\{ f \in \mathcal{F} \ \Bigg| \lim_{\delta
\rightarrow 0} \int_{0}^{1} \frac{|f(x) - f(x - \delta)|}{|\delta|}
dx < +\infty \right\}.
\]
A function in $\mathcal{F}_{BV}$ has a derivative (at points for
which it exists) which is uniformly bounded and the sum of the
amplitudes of its discontinuous jumps is finite. The
bounded--variation condition represents a minimal ``smoothness''
assumption since a restriction is placed on the amount of total
discontinuous jumps.

It is well known that for the Fourier basis,
\begin{equation} \label{eqn:FourierBasis}
\mathcal{B}_{\mathrm{Fourier}} = \left\{ \phi_j(x) = \begin{cases}
e^{+\pi j x \sqrt{-1}}, & j~\mbox{even}, \\
e^{-\pi (j+1) x \sqrt{-1}}, &
j~\mbox{odd}
\end{cases} \right\}_{j=0}^{\infty},
\end{equation}
the $m$--term approximation error \eqnref{eqn:mTermError} is upper
bounded as follows,
\begin{equation} \label{eqn:varepsBV}
\forall f \in \mathcal{F}_{BV}, \quad
\varepsilon[f,m,\mathcal{B}_{\mathrm{Fourier}}] \leq
\frac{\sigma}{m},
\end{equation}
where $\sigma > 0$ is a constant
\cite[Chapter~9]{Mallat-AP98-WToSP}. Combining
\eqnref{eqn:MSEUpperBound} with \eqnref{eqn:varepsBV} yields the
following corollary.

{\bf
\begin{corollary} \label{thm:MSERateBV}
{\it \textbf{(Decay rate of integrated MSE for $\mathcal{F}_{BV}$)}
Let $\mathcal{F}_{BV}$ be as given above and $\mathcal{P}_Z$ and $p_X$
be as given in \secref{sec:problemSetup}. Let $\hat{f}_{n,m}$ be given
by \eqnref{eqn:coefEst} and \eqnref{eqn:finiteMEstFunc} with
$\mathcal{B}_{\mathrm{Fourier}}$ as given in
\eqnref{eqn:FourierBasis}. If $p_X$ satisfies \eqnref{eqn:deployCond},
then $\forall f \in \mathcal{F}_{BV}$ and $\forall p_Z \in
\mathcal{P}_Z$, the integrated MSE of $\hat{f}_{n,m}$ is upper bounded
as follows
\[
D = \eE \left[ \|f - \hat{f}_{n,m}\|^2 \right] \leq \frac{c^2 m}{n
\nu} + \frac{\sigma}{m},
\]
where $\sigma > 0$ is a constant. Setting $m(n) = \sqrt{n}$ to
optimize the decay rate of the upper bound yields the following
achievable upper rate of convergence of the integrated MSE $\forall f
\in \mathcal{F}_{BV}$ and $\forall p_Z \in \mathcal{P}_Z$:
\[
D = \eE \left[ \|f - \hat{f}_{n,m}\|^2 \right] = O
\left(\frac{1}{\sqrt{n}} \right).
\]}
\end{corollary}
}

\subsection{Sobolev differentiable functions on Domain $\mathcal{D} = [0,1]$} \label{sec:smoothFuncClass}
This function class includes functions which are differentiable in a
generalized sense to a degree of differentiability parameterized by
$s$ which can take non--integer values. The value of $s$ can be
considered as a measure of smoothness. For $s>1/2$, let
$\mathcal{F}_s$ denote the subset of $\mathcal{F}$ which is composed
of functions on $\mathcal{D} = [0,1]$ that are $s$--times Sobolev
differentiable. Formally,
\begin{equation} \label{eqn:sobolevCond}
\mathcal{F}_s := \left\{ f \in \mathcal{F} \ \Bigg|
\int_{-\infty}^{+\infty} |\omega|^{2s} |\tilde{f}(\omega)|^2 d\omega
< +\infty \right\},
\end{equation}
where $\tilde{f}(\omega)$ denotes the Fourier transform of $f$. Note
that the condition in \eqnref{eqn:sobolevCond} (for integer values
of $s$) corresponds to the $s^{\mathrm{th}}$ derivative of $f$
belonging to $\mathbf{L}^2([0,1])$. Thus, this set includes
functions that are $\lfloor s \rfloor$--times differentiable.

It is well known that for $s > 1/2$,
\begin{equation} \label{eqn:varepsSmooth}
\forall f \in \mathcal{F}_s, \quad
\varepsilon[f,m,\mathcal{B}_{\mathrm{Fourier}}] \leq
\frac{\sigma}{m^{2s}},
\end{equation}
where $\sigma > 0$ is a constant
\cite[Chapter~9]{Mallat-AP98-WToSP}. Combining
\eqnref{eqn:MSEUpperBound} with \eqnref{eqn:varepsSmooth} yields the
following corollary.

{\bf
\begin{corollary} \label{thm:MSERateSmooth}
{\it \textbf{(Decay rate of integrated MSE for $\mathcal{F}_s$)} Let
$\mathcal{F}_s$ be as given above and $\mathcal{P}_Z$ and $p_X$ be as
given in \secref{sec:problemSetup}. Let $\hat{f}_{n,m}$ be given by
\eqnref{eqn:coefEst} and \eqnref{eqn:finiteMEstFunc} with
$\mathcal{B}_{\mathrm{Fourier}}$ as given in
\eqnref{eqn:FourierBasis}. If $p_X$ satisfies \eqnref{eqn:deployCond},
then $\forall f \in \mathcal{F}_s$ and $\forall p_Z \in
\mathcal{P}_Z$, the integrated MSE of $\hat{f}_{n,m}$ is upper bounded
as follows
\[
D = \eE \left[ \|f - \hat{f}_{n,m}\|^2 \right] \leq \frac{c^2 m}{n
\nu} + \frac{\sigma}{m^{2s}},
\]
where $\sigma > 0$ is a constant. Setting $m(n) = n^{\frac{1}{2s +
1}}$ to optimize the decay rate of the upper bound yields the
following achievable upper rate of convergence of the integrated MSE $
\forall f \in \mathcal{F}_s$ and $\forall p_Z \in \mathcal{P}_Z$:
\[
D = \eE \left[ \|f - \hat{f}_{n,m}\|^2 \right] = O
\left(n^{\frac{-2s}{2s + 1}} \right).
\]}
\end{corollary}
}

It is well known that the exact minimax rate of convergence of the
integrated MSE for non--parametric regression, based on
full--resolution, real--valued, noisy observations in an
$s$--Sobolev space is given by $\gamma_n = n^\frac{-2s}{2s + 1}$
\cite{KorostelevT-Springer93-MToIR}, \cite{GolubevN-AoS1990-RBSCR}.
In non--parametric regression, the field estimate is based directly
on the full--resolution real--valued noisy observations
$\{Y_i\}_{i=1}^n$, whereas in our problem the field estimate is
based on only the binary--quantized observations $\{B_i\}_{i=1}^n$.
In both setups, the corresponding sensor locations are known. Thus,
it is interested to observe that our proposed estimator is minimax
order optimal even with respect to the case in which the
observations have not been quantized.

\section{Proofs} \label{sec:proofs}

\subsection{Proof of \thmref{thm:MSEUpperBound} \label{sec:MainProof}}
We first establish some results regarding the estimated coefficients
of \eqnref{eqn:coefEst}.

{\bf
\begin{lemma} \label{lem:coefProps}
{\it The expected value of an approximated coefficient is given by
\begin{equation} \label{eqn:alphaExptVal}
\mbox{(i) } \eE[\hat{\alpha}_j] = \alpha_j = \langle f,\phi_j
\rangle,
\end{equation}
and the integrated MSE of the coefficient estimates satisfies
\begin{equation} \label{eqn:alphaVarBound}
\mbox{(ii) }\eE[|\hat{\alpha}_j - \alpha_j|^2] \leq \frac{c^2}{n}
\int_{\mathcal{D}} \frac{|\phi_j(x)|^2}{p_X(x)} dx.
\end{equation}
The approximated coefficients also have the following convergence
property
\begin{equation} \label{eqn:alphaASConv}
\mbox{(iii) }\hat{\alpha}_j \xrightarrow{\mathrm{a.s.}} \alpha_j,
\quad n \longrightarrow \infty.
\end{equation} }
\end{lemma}
}

\begin{proof}
(i) The expectation of the coefficient estimates can be evaluated as
follows
\begin{eqnarray} \label{eqn:alphaExptEval}
\eE[\hat{\alpha}_j] &=& \eE\left[\frac{c}{n} \sum_{i=1}^n
\frac{\phi_j^*(X_i)}{p_X(X_i)} B_i \right]
\nonumber \\
&=& \frac{c}{n} \sum_{i=1}^n \eE\left[\frac{\phi_j^*(X_i)}{p_X(X_i)}
\mathrm{sgn}
(f(X_i) + Z_i - T_i) \right] \nonumber \\
&=& c\eE\left[\frac{\phi_j^*(X_1)}{p_X(X_1)}\mathrm{sgn} (f(X_1) +
Z_1 - T_1)\right],
\end{eqnarray}
where the last equality follows since the terms are iid. This last
expectation can be evaluated as follows
\begin{align} \label{eqn:auxExptEval}
& \eE\left[\frac{\phi_j^*(X_1)}{p_X(X_1)}\mathrm{sgn}
(f(X_1) + Z_1 - T_1)\right] \nonumber \\
& = \int_{\mathcal{D}} \hspace{-5pt} p_X(x) \int_{-b}^{+b}
\hspace{-10pt} p_Z(z) \int_{-c}^{+c} \! \frac{1}{2c} \frac{\phi_j^*(x)}{p_X(x)} \mathrm{sgn}(f(x) + z - t) dt dz dx \nonumber \\
& = \int_{\mathcal{D}} \int_{-b}^{+b} p_Z(z) \phi_j^*(x)\frac{1}{2c}
\left( \int_{-c}^{f(x)+z} \!\!\!\!\! dt - \int_{f(x)+z}^{+c} \!\!\!
dt
\right) dz dx \nonumber \\
& = \frac{1}{c} \int_{\mathcal{D}} \int_{-b}^{+b} p_Z(z) \phi_j^*(x)
(f(x)
+ z) dz dx \nonumber \\
& = \frac{1}{c} \int_{\mathcal{D}} \phi_j^*(x) f(x) dx \nonumber \\
& = \frac{1}{c}\langle f,\phi_j \rangle = \frac{\alpha_j}{c},
\end{align}
where the second to last line follows from the assumption that $p_Z$
is a zero--mean distribution. Combining \eqnref{eqn:alphaExptEval}
and \eqnref{eqn:auxExptEval}, we have \eqnref{eqn:alphaExptVal}.
(ii) Thus,
\begin{equation}\label{eqn:alphaVar1}
\eE[|\hat{\alpha}_j - \alpha_j|^2] = \eE[|\hat{\alpha}_j -
\eE[\hat{\alpha}_j]|^2] = \mathrm{Var}[\hat{\alpha}_j].
\end{equation}
Using standard properties of variance and the fact that the terms
$\{\phi_j^*(X_i)B_i/p_X(X_i)\}_{i=1}^n$ are iid, we obtain the
following
\begin{eqnarray} \label{eqn:alphaVar2}
\mathrm{Var}[\hat{\alpha}_j] &=& \frac{c^2}{n^2}\sum_{i=1}^{n}
\mathrm{Var}\left[\frac{\phi_j^*(X_i)}{p_X(X_i)} B_i\right] \nonumber \\
&=& \frac{c^2}{n} \mathrm{Var}\left[\frac{\phi_j^*(X_1)}{p_X(X_1)}
B_1\right] \nonumber \\
&=& \frac{c^2}{n} \eE\left[\left|\frac{\phi_j^*(X_1)}{p_X(X_1)}
B_1\right|^2\right] - \frac{c^2}{n}
\left|\eE\left[\frac{\phi_j^*(X_1)}{p_X(X_1)}
B_1\right]\right|^2 \nonumber \\
&\leq& \frac{c^2}{n}
\eE\left[\left|\frac{\phi_j^*(X_1)}{p_X(X_1)}\right|^2\right], \nonumber \\
&=& \frac{c^2}{n} \int_{\mathcal{D}} \frac{|\phi_j(x)|^2}{p_X^2(x)}
p_X(x) dx, \nonumber \\
&=& \frac{c^2}{n} \int_{\mathcal{D}} \frac{|\phi_j(x)|^2}{p_X(x)}
dx.
\end{eqnarray}
Combining \eqnref{eqn:alphaVar1} and \eqnref{eqn:alphaVar2}, we
arrive at \eqnref{eqn:alphaVarBound}. (iii) The coefficient
estimates
\begin{eqnarray} \label{eqn:alphaASInter}
\hat{\alpha}_j &=& \frac{c}{n} \sum_{i=1}^n
\frac{\phi_j^*(X_i)}{p_X(X_i)} B_i
\nonumber \\
&=& \frac{c}{n} \sum_{i=1}^n \frac{\phi_j^*(X_i)}{p_X(X_i)}
\mathrm{sgn}
(f(X_i) + Z_i - T_i) \nonumber \\
&\xrightarrow{\mathrm{a.s.}}&
c\eE\left[\frac{\phi_j^*(X_1)}{p_X(X_1)}\mathrm{sgn} (f(X_1) + Z_1 -
T_1)\right],
\end{eqnarray}
as $n \longrightarrow \infty$ by Kolmogorov's strong law of large
numbers since each term in the summation is iid and has a first
moment bounded by $\sqrt{\mathrm{vol}(\mathcal{D})}$:
\begin{eqnarray*}
\eE \left[ \left| \frac{\phi_j(X_1)}{p_X(X_1)} \right| \right] = \|
\phi_j\|_1 \leq \sqrt{\mathrm{vol}(\mathcal{D})} \|\phi_j \|_2 =
\sqrt{\mathrm{vol}(\mathcal{D})},
\end{eqnarray*}
where the last inequality follows from the Cauchy-–Schwartz
inequality.  Combining \eqnref{eqn:auxExptEval} and
\eqnref{eqn:alphaASInter}, we obtain \eqnref{eqn:alphaASConv},
concluding the proof of the \lemref{lem:coefProps}.
\end{proof}

For any orthonormal basis $\mathcal{B} = \{\phi_j\}_{j=0}^{\infty}$
and for any field $f \in \mathcal{F}$, the integrated MSE of the
estimate can be written as follows
\begin{eqnarray} \label{eqn:generalMSE}
D &=& \eE[\|f - \hat{f}_{n,m}\|^2] \nonumber \\
&=& \eE\left[\left\|\sum_{j=0}^{\infty} \alpha_j \phi_j -
\sum_{j=0}^{m-1} \hat{\alpha}_j \phi_j\right\|^2\right] \nonumber \\
&=& \sum_{j=0}^{m-1} \eE[|\hat{\alpha}_j - \alpha_j|^2] +
\sum_{j=m}^{\infty} |\alpha_j|^2 \nonumber \\
&\leq& \frac{c^2}{n} \sum_{j=0}^{m-1} \int_{\mathcal{D}}
\frac{|\phi_j(x)|^2}{p_X(x)} dx + \underbrace{\sum_{j=m}^{\infty}
|\alpha_j|^2}_{=\varepsilon[f,m,\mathcal{B}]},
\end{eqnarray}
where in the last step we used the bound given in
\eqnref{eqn:alphaVarBound}. Thus we have \eqnref{eqn:MSEUpperBound},
concluding the proof of \thmref{thm:MSEUpperBound}.
\QED

\subsection{Proof of \thmref{thm:ErrorASConv} \label{sec:AsErrorProof}}
The pointwise errors of the field estimate with respect to the
$m$--term approximation can be written as
\begin{eqnarray*}
S_n(x) &:=& \hat{f}_{n}(x) - f_{m(n)}(x) \\
&=& \sum_{j=0}^{m(n)-1} \left( \frac{1}{n} \sum_{i=1}^{n} \frac{c
\phi_j^*(X_i)B_i}{p_X(X_i)} - \alpha_j \right) \phi_j(x) \\
&=& \frac{1}{n} \sum_{i=1}^{n} \sum_{j=0}^{m(n)-1} \left(\frac{c
\phi_j^*(X_i)B_i}{p_X(X_i)} - \alpha_j \right) \phi_j(x) \\
&=& \frac{1}{n} \sum_{i=1}^{n} U_i(x),
\end{eqnarray*}
where for $i \in \{1,\ldots, n\}$,
\[
U_i(x) := \sum_{j=0}^{m(n)-1} \left(\frac{c
\phi_j^*(X_i)B_i}{p_X(X_i)} - \alpha_j \right) \phi_j(x).
\]
Note that $U_i(x)$ is iid and that it is zero--mean due to
\eqnref{eqn:alphaExptVal} of \lemref{lem:coefProps}. However, almost
sure convergence of $S_n(x)$ cannot be directly deduced from the
standard strong law of large numbers since the distribution of
$U_i(x)$ itself depends on $n$ because it is the summation of $m(n)$
terms. Instead, we leverage a more fundamental condition for almost
sure convergence \cite[p.~206]{Ash-Wiley70-BPT}: if for all
$\epsilon > 0$,
\begin{equation*}
\sum_{n=1}^\infty \pP[|S_n(x)| \geq \epsilon] < \infty,
\end{equation*}
then $S_n(x) \xrightarrow{\mathrm{a.s.}} 0$ as $n \longrightarrow
\infty$.

Associated with $S_n(x)$, is a martingale $\{V_k(x)\}_{k=0}^{n}$ given
by $V_0 := 0$, and for $k \in \{1,\ldots, n\}$,
\[
V_k(x) := \sum_{i=1}^{k} \frac{1}{n} U_i(x).
\]
$V_0(x), \ldots, V_n(x)$ is a martingale since
$\{U_i(x)\}_{i=1}^{n}$ is iid with zero--mean. Note that $V_n(x) =
S_n(x)$ and that $|V_k(x) - V_{k-1}(x)| \leq |\frac{1}{n} U_k(x)|$.
For each $i \in \{1, \ldots, n\}$,
\[
|U_i(x)| \leq c \left| \sum_{j=0}^{m(n)-1}
\frac{\phi_j(x)\phi_j^*(X_i)}{p_X(X_i)} \right| + \left|
\sum_{j=0}^{m(n)-1} \alpha_j \phi_j(x) \right|,
\]
by the triangle inequality. Under the assumptions that the
conditions given by \eqnref{eqn:ASConvCond1} and
\eqnref{eqn:ASConvCond2} hold and that the deployment distribution
$p_X$ is non--singular, there exists some constant $C > 0$ such that
for all $i \in \{1, \ldots, n\}$,
\[
|U_i(x)| \leq C \Lambda_{m(n)},
\]
with probability $1$ for $x \in \mathcal{D}$ almost everywhere. Thus
\begin{equation} \label{eqn:ASPrfAuxRVBnd}
|V_k(x) - V_{k-1}(x)| \leq \frac{C \Lambda_{m(n)}}{n}.
\end{equation}
According to the Azuma--Hoeffding inequality (see
\cite[p.~303]{MitzenmacherU-Cambrdge05-PaC}), if for all $k \in
\{1,\ldots, n\}$, $|V_k(x) - V_{k-1}(x)| \leq C_k$, then for all
$\epsilon > 0$,
\begin{eqnarray*}
\pP[|V_n(x)| \geq \epsilon] &\leq& 2\exp\left(\frac{-\epsilon^2}{2
\sum_{k=1}^n C_k^2} \right) \nonumber.
\end{eqnarray*}
Applying this inequality with $C_k = C\Lambda_{m(n)}/n$ for all $k
\in \{1,\ldots, n\}$ (see \eqnref{eqn:ASPrfAuxRVBnd}) and $V_n(x) =
S_n(x)$ we obtain the following upper bound
\begin{eqnarray*}
\pP[|S_n(x)| \geq \epsilon] &\leq& 2\exp\left(\frac{-\epsilon^2}{2
\sum_{k=1}^n \frac{C^2 \Lambda_{m(n)}^2}{n^2}} \right) \nonumber \\
&=& 2\exp\left(\frac{-\epsilon^2n}{2 C^2 \Lambda_{m(n)}^2} \right),
\end{eqnarray*}
for $x \in \mathcal{D}$ almost everywhere. Therefore,
\[
\sum_{n=1}^\infty \pP[|S_n(x)| \geq \epsilon] \leq \sum_{n=1}^\infty
2\exp\left(\frac{-\epsilon^2n}{2 C^2 \Lambda_{m(n)}^2} \right),
\]
which is less than infinity $\forall \epsilon > 0$ and $x \in
\mathcal{D}$ almost everywhere, due to condition
\eqnref{eqn:ASConvCond3}. Thus, as $n \longrightarrow \infty$,
almost surely $S_n(x) \longrightarrow 0$, for $x \in \mathcal{D}$
almost everywhere. \QED

\subsection{Proof of \prpref{prp:asAuxFcnsPrp} \label{sec:PropProof}}

Part (i): Note that if $|g_{\infty}(x,y)| = 0$ for $x, y \in
\mathcal{D}$ almost everywhere and $|h_{\infty}(x)| = 0$ for all $f
\in \mathcal{F}$ and for $x \in \mathcal{D}$ almost everywhere, then
conditions \eqnref{eqn:ASConvCond1} and \eqnref{eqn:ASConvCond2}
hold with some constants $C_1, C_2 > 0$. For $g_m$, we can write
\begin{eqnarray*}
\lefteqn{
\iint_{\mathcal{D} \times \mathcal{D}} p_X^2(y) |g_{\infty}(x,y)|^2
dx dy = } \\
&=& \iint_{\mathcal{D} \times \mathcal{D}} \liminf_{m
\longrightarrow \infty} p_X^2(y) |g_m(x,y)|^2 dx dy \\
&\stackrel{\mathrm{(a)}}{\leq}& \liminf_{m \longrightarrow \infty}
\iint_{\mathcal{D} \times
\mathcal{D}} p_X^2(y) |g_m(x,y)|^2 dx dy \\
&=& \liminf_{m \longrightarrow \infty} \iint_{\mathcal{D} \times
\mathcal{D}} p_X^2(y) \frac{1}{\Lambda_m^2} \sum_{j=0}^{m-1}
\sum_{k=0}^{m-1} \phi_j(x)
\phi_j^*(y) \cdot \\
&& \hspace{100pt} \phi_k^*(x) \phi_k(y) \frac{1}{p_X^2(y)} dx dy \\
&=& \liminf_{m \longrightarrow \infty} \frac{1}{\Lambda_m^2}
\sum_{j=0}^{m-1} \sum_{k=0}^{m-1} \underbrace{\int_{\mathcal{D}}
\phi_j(x) \phi_k^*(x) dx}_{= \delta_{j-k}} \cdot \\
&& \hspace{100pt} \underbrace{\int_{\mathcal{D}} \phi_k(y)
\phi_j^*(y) dy}_{= \delta_{j-k}} \\
&=& \liminf_{m \longrightarrow \infty} \frac{1}{\Lambda_m^2}
\sum_{j=0}^{m-1} 1 = \liminf_{m \longrightarrow \infty}
\frac{m}{\Lambda_m^2},
\end{eqnarray*}
where the inequality (a) is due to Fatou's lemma
\cite{Rudin-McGraw74-RaCA} and $\delta_k$ is the Kronecker delta
function. Thus for $\Lambda_m$ such that \eqnref{eqn:ASBoundCond} is
satisfied we have that
\[
\iint_{\mathcal{D} \times \mathcal{D}} p_X^2(y) |g_{\infty}(x,y)|^2
dx dy = 0,
\]
which implies that $|g_{\infty}(x,y)| = 0$ for $x, y \in
\mathcal{D}$ almost everywhere due to \eqnref{eqn:deployCond}. For
$h_m$, we can write
\begin{eqnarray*}
\lefteqn{\int_{\mathcal{D}} |h_{\infty}(x)|^2 dx = \int_{\mathcal{D}}
\liminf_{m \longrightarrow \infty} |h_m(x)|^2 dx } \\
&\leq& \liminf_{m \longrightarrow \infty} \int_{\mathcal{D}}
|h_m(x)|^2 dx\\
&=& \liminf_{m \longrightarrow \infty} \int_{\mathcal{D}}
\frac{1}{\Lambda_m^2} \sum_{j=0}^{m-1} \sum_{k=0}^{m-1} \alpha_j
\alpha_k^* \phi_j(x)
\phi_k^*(x) dx \\
&=& \liminf_{m \longrightarrow \infty} \frac{1}{\Lambda_m^2}
\sum_{j=0}^{m-1} \sum_{k=0}^{m-1} \alpha_j \alpha_k^*
\underbrace{\int_{\mathcal{D}}
\phi_j(x) \phi_k^*(x) dx}_{= \delta_{j-k}} \\
&=& \liminf_{m \longrightarrow \infty} \frac{1}{\Lambda_m^2}
\sum_{j=0}^{m-1} |\alpha_j|^2 \\
&\leq& \liminf_{m \longrightarrow \infty} \frac{\|f\|^2}{\Lambda_m^2} \\
&\leq& \liminf_{m \longrightarrow \infty} \frac{a^2
\mathrm{vol}(\mathcal{D})}{\Lambda_m^2}, \quad \forall f \in
\mathcal{F},
\end{eqnarray*}
where the first inequality follows from Fatou's lemma
\cite{Rudin-McGraw74-RaCA} and the last inequality is due to $f$
being amplitude--bounded by $a$ over the support $\mathcal{D}$. Thus
for $\Lambda_m$ such that \eqnref{eqn:ASBoundCond} is satisfied, we
have that
\[
\forall f \in \mathcal{F}, \quad \int_{\mathcal{D}} |h_{\infty}(x)|^2
dx = 0,
\]
which implies that $|h_{\infty}(x)| = 0$ for all $f \in \mathcal{F}$
and for $x \in \mathcal{D}$ almost everywhere.

Part (ii): Applying the triangle inequality, we can write
\begin{align*}
&\left|\sum_{j=0}^{m-1} \phi_j(x) \phi_j^*(y)\frac{1}{p_X(y)}\right| \\
&\leq \sum_{j=0}^{m-1} \frac{|\phi_j(x)| |\phi_j^*(y)|}{p_X(y)} \\
&\leq \sum_{j=0}^{m-1} \frac{\beta^2}{\nu} = \frac{\beta^2}{\nu} m =
C_1 \Lambda_m,
\end{align*}
which shows that condition \eqnref{eqn:ASConvCond1} is satisfied for
$\Lambda_m = m$ and $C_1 = \beta^2/\nu$. Again, applying the
triangle inequality, we can write
\begin{align*}
& \left|\sum_{j=0}^{m-1} \langle f,\phi_j \rangle \phi_j(x) \right|\\
& \leq \sum_{j=0}^{m-1} |\langle f,\phi_j \rangle| |\phi_j(x)| \\
& \leq \sum_{j=0}^{m-1} \|f\| \beta  = m \|f\| \beta \\
& \leq m a \beta \sqrt{\mathrm{vol}(\mathcal{D})} = C_2 \Lambda_m,
\end{align*}
which shows that condition \eqnref{eqn:ASConvCond2} is satisfied for
$\Lambda_m = m$ and $C_2 = a \beta
\sqrt{\mathrm{vol}(\mathcal{D})}$. \QED

\section{Concluding Remarks} \label{sec:concl}
The principal contribution of this work is a systematic treatment of
(i) binary--{\em sensing}, (ii) random sensor {\em deployment}, and
(iii) unknown observation noise distribution for high--resolution
distributed sensing and estimation of multidimensional fields using
dense sensor networks. A key finding of this work is that the rate
of convergence of the integrated MSE for field estimation is
extremely robust to the apparent limitations of ultra--poor sensing
precision, random sensor deployment, and lack of knowledge of
observation noise statistics. In some cases, the convergence rate
exactly matches the minimax rate of convergence with
infinite--precision real--valued samples and known noise statistics.
Interesting directions for future work include (i) establishing the
exact rate of convergence of the integrated MSE and a central limit
theorem for the estimate, (ii) analysis of the sensitivity of the
integrated--MSE to sensor location uncertainty, (iii)
unbounded--amplitude signal and noise models, and (iv) general
dither distributions.

\section*{Acknowledgment}
The authors would like to thank Professor Elias Masry, Department of
Electrical and Computer Engineering at the University of California
San Diego, for his encouraging comments.

\bibliographystyle{IEEEtran}
\bibliography{../LaTeX/biblio}

\end{document}